\documentclass[prd,twocolumn,preprintnumbers,floatfix,superscriptaddress,nofootinbib,amsmath,amssymb]{revtex4}

\usepackage{graphicx}
\usepackage{epsfig}
\usepackage{bbm}

\def\bfx{{\bf x}}

\newcommand{\FIGURE}[1]{\
\begin{figure}[!tb]
#1
\end{figure}
}

\newcommand{\TABLE}[1]{\
\begin{table}[!tb]
#1
\end{table}
}
\newcommand{\WTABLE}[1]{\
\begin{table*}[!tb]
#1
\end{table*}
}

\begin{document}

\title{Higher representations on the lattice: Numerical simulations.\\
SU(2) with adjoint fermions.
}

\author{Luigi Del Debbio}
\affiliation{SUPA, School of Physics and Astronomy, University of Edinburgh, Edinburgh EH9 3JZ, Scotland, UK}
\email[E-mail: ]{luigi.del.debbio@ed.ac.uk} 
\author{Agostino Patella}
\affiliation{School of Physics, Swansea University, Swansea SA2 8PP, Wales, UK}
\email[E-mail: ]{a.patella@swan.ac.uk} 
\author{Claudio Pica}
\affiliation{CP$^\mathit{3}$-Origins, University of Southern Denmark, 5230, Odense M, Denmark}
\email[E-mail: ]{pica@cp3.sdu.dk} 
\affiliation{Physics Department, Brookhaven National Laboratory, Upton, NY 11973-5000, USA}

\begin{abstract}
	We discuss the lattice formulation of gauge theories with
	fermions in arbitrary representations of the color group, and
	present in detail the implementation of the HMC/RHMC algorithm for
	simulating dynamical fermions. We discuss the validation of the
	implementation through an extensive set of tests, and the stability
	of simulations by monitoring the distribution of the lowest
	eigenvalue of the Wilson-Dirac operator.
	Working with two flavors of Wilson fermions in the adjoint
        representation, benchmark results for realistic lattice
        simulations are presented. Runs are performed on different
        lattice sizes ranging from $4^3 \times 8$ to $24^3\times 64$
        sites. For the two smallest lattices we also report the
        measured values of benchmark mesonic observables. These
        results can be used as a baseline for rapid cross-checks of
        simulations in higher representations.
	The results presented here are the first steps towards more
	extensive investigations with controlled systematic errors, aiming
	at a detailed understanding of the phase structure of these
	theories, and of their viability as candidates for strong dynamics
	beyond the Standard model.
\end{abstract}


\preprint{NI08019, BNL-NT-08/15, CP3-Origins-2010-15}

\maketitle

\section{Introduction}
Recent algorithmic progress in simulations of QCD with dynamical
fermions have shown that current computing facilities can reach beyond
the quenched approximation without compromising the robustness of the
formulation~\cite{Hasenbusch:2001ne,Hasenbusch:2002ai,Luscher:2005rx,DelDebbio:2005qa,Urbach:2005ji,Clark:2006fx}. Besides
their great relevance for lattice QCD, these improvements allow
numerical simulations to be used as a quantitative, nonperturbative
tool to study more generic strongly--interacting theories where the
role of the fermion determinant is crucial: extensive Monte Carlo
studies will enable to tame the systematic errors and therefore to
obtain reliable information about the strong dynamics of theories
beyond QCD. Examples that have a direct impact on physics beyond the
Standard Model (BSM) include technicolor
models~\cite{Weinberg:1975gm,Susskind:1978ms}
(see~\cite{Hill:2002ap,Sannino:2008ha} for reviews), orientifold
theories~\cite{Armoni:2003gp}, and supersymmetry (see
\cite{Kaplan:2007zz} for a recent review on the nonperturbative
formulation of supersymmetry). Moreover the theoretical progresses in
formulating QCD on the lattice provide the necessary tools for
rigorous studies of BSM nonperturbative phenomena.  This is of
paramount importance in order to obtain reliable results as we move
into the unknown territories of strongly--interacting BSM theories:
the clean field--theoretical framework developed for QCD needs to be
exported to the studies of BSM strong dynamics in order to obtain
lattice results that are trustable and therefore can have an impact on
the LHC phenomenology.

Simulations of the theories listed above share a common technical
challenge: they all require fermion fields in representations larger
than the fundamental, gauge groups SU($N$) with generic number of
colors $N$, or a combination of both. Therefore a central ingredient
of these simulations is a generalized Dirac operator, that can act on
vectors of arbitrary dimension in color space. We developed such an
operator and tested its inversion in a recent study of the quenched
meson spectrum in the large-$N$ limit of pure Yang--Mills
theories~\cite{DelDebbio:2007wk}. The same computational framework has
also been used to assess the validity of the planar orientifold
equivalence by numerical simulations in the quenched
approximation~\cite{Armoni:2008nq}.

Theories with two Dirac fermions in the two--index symmetric
representation and small number of colors ($N=2,3$) are currently
under scrutiny as viable candidates for strongly--interacting BSM
candidates~\cite{Sannino:2004qp,Dietrich:2006cm,Foadi:2007ue}. An
important feature of these theories is that the number of fermions is
close to the critical value where an IR conformal fixed point may
appear~\cite{Banks:1981nn}. The vicinity of the IR fixed point can
induce a walking behaviour of the running coupling, which would make
such theories phenomenologically viable. Whether or not this is really
the case beyond perturbation theory can only be established by
numerical simulations of such theories defined on the lattice. Recent
studies of the running of the renormalized coupling in the
Schr\"odinger functional (SF) scheme~\cite{Luscher:1992an} with
fundamental staggered fermions have provided first evidence of the
existence of a conformal fixed point for
$n_f=12$~\cite{Appelquist:2007hu}. A similar study has also appeared
for the SU(3) theory with fermions in the two--index symmetric
representation~\cite{Shamir:2008pb}, and for the SU(2) theory with
fermions in the adjoint
representation~\cite{Hietanen:2009az,Bursa:2009we}. The results
presented in those works confirm that the SF is indeed a valuable tool
to study the RG structure of the theories under consideration. Further
studies on the existence of an IR fixed point with fundamental
fermions have appeared in the
literature~\cite{Deuzeman:2008sc}. Several complementary approaches
will be needed in order to fully understand numerical simulations in
the vicinity of an IR fixed point.

In this paper we discuss in detail the implementation of an algorithm
for simulating gauge theories with dynamical Wilson fermions in an
arbitrary representation, and arbitrary gauge group SU($N$), and
present the results of preliminary runs which are useful as a benchmark for future simulations.
 At this stage in our
investigations, the emphasis is on the algorithm rather than the
optimization or the phenomenological results. Sect.~\ref{sec:dirac}
summarizes the notation used for the Dirac operator, and introduces
the conventions to deal with the higher representations. The HMC
algorithm for two flavors is described in Sect.~\ref{sec:hmc}. In
order to be able to simulate theories with an arbitrary number of
flavors, the RHMC algorithm~\cite{Clark:2006fx} has also been
developed, as described in Sect.~\ref{sec:rhmc}. Let us emphasize once
again that, as numerical simulations move away from the well--known
realms of QCD, we find it necessary to have a robust
field--theoretical formulation and a tight control over the behaviour
of the numerical simulations. These are necessary conditions if
numerical simulations want to be in a position to test candidate
theories for strong electroweak symmetry breaking with any degree of
confidence.


We study the behaviour of the algorithm at different points in the
space of bare lattice couplings by monitoring the time--history, the
probability distribution and the correlation time of the plaquette and
the lowest eigenvalue of the Dirac operator, in
Sect.~\ref{sec:res}. In the same Section we also study how the force
is split in the terms of the rational approximation, in analogy with
Ref.~\cite{Clark:2006fx} where fundamental fermions were
considered. We present results for $4^3\times 8$, $8^3\times 16$,
$12^3\times 24$, $16^3\times 32$, $24^3\times 64$ lattices, for the
SU(2) theory with two flavours in the adjoint representation at two
values of the coupling constant $\beta=2$ and $\beta=2.25$. The
emphasis in this work is on the details of the implementation, and on
the behaviour of the algorithm for the theory under
consideration. These results are intended as benchmark for other
softwares that simulate the same theory. When the exactness of
numerical simulations cannot be judged by comparing with experimental
data, a cross--check of all the available simulation softwares is
essential before debating about the interpretation of the results. Up
to date this preliminary cross--check is still missing, although
several more complex and phenomenologically relevant results are
currently available in the
literature~\cite{Catterall:2007yx, Catterall:2009sb, Catterall:2008qk, Hietanen:2009zz, Hietanen:2009az, Hietanen:2008mr, Bursa:2009tj, Bursa:2009we}
(for studies with different number of colors or fermionic representation, see
\cite{Deuzeman:2008sc, Deuzeman:2009mh, Fodor:2009ar, Fodor:2009wk, Fodor:2009rb, Appelquist:2007hu, Appelquist:2009ty, Appelquist:2009ka, Appelquist:2010sc, Jin:2009mc, Shamir:2008pb, DeGrand:2008kx, DeGrand:2009mt, DeGrand:2009et, DeGrand:2009hu, Svetitsky:2009pz, Machtey:2009wu, Hasenfratz:2009ea, Hasenfratz:2009kz, Sinclair:2009ec, Kogut:2010cz,Bilgici:2009nm}). Therefore
the authors encourage different groups to publish analogous studies,
in order to fill this gap.

Some phenomenological results for the simulations presented in this
work have been already published in a short
paper~\cite{DelDebbio:2009fd} (see also~\cite{Pica:2009hc,Lucini:2009an}). A more detailed discussion of these
results is currently in preparation.

\section{Wilson fermions in a generic representation}
\label{sec:dirac}
In four--dimensional Euclidean space the massless Wilson--Dirac
operator is written following the notation in
Ref.~\cite{Luscher:1996sc}:
\begin{equation}
D = \frac12 \sum_\mu \left[\gamma_\mu \left(\nabla_\mu + \nabla^*_\mu \right) 
- a \nabla^*_\mu \nabla_\mu \right],
\end{equation}
where $\nabla$ and $\nabla^*$ indicate respectively the forward and
backward covariant derivatives:
\begin{eqnarray}
  \label{eq:covdev}
  \nabla_\mu f(x) &=& \frac{1}{a}\left[U(x,\mu) f(x+\mu) - f(x)\right]\, , \\
  \nabla^*_\mu f(x) &=& \frac{1}{a}\left[f(x) - U(x-\mu,\mu)^\dagger f(x-\mu)\right]\, ,
\end{eqnarray}
and $U(x,\mu)$ denote the link variables as usual.  This expression is
easily generalized to an arbitrary representation $R$ of the gauge
group; the action of the massive Dirac operator on a spinor field
$\psi$ yields:
\begin{eqnarray}
D_m \psi(x) &\equiv& (D+m_0) \psi(x) =\nonumber \\
&=& (4/a+m_0)\psi(x) - \nonumber\\
&& - \frac{1}{2a} \sum_\mu \left\{ \left(1-\gamma_\mu\right)\times U^R(x,\mu)\vphantom{^\dagger}
  \psi(x+\mu) + \right. \nonumber \\
&& + \left. \left(1+\gamma_\mu\right) U^R(x-\mu,\mu)^\dagger \psi(x-\mu) \right\} \label{eq:Dirac}, 
\end{eqnarray}
where $U^R$ are the link variables in the representation $R$, and $m_0$
is the bare mass. 


Let $T_F^a,T_R^a$ be respectively the generators in the fundamental
representation and in a generic representation $R$. Our conventions
for the normalization of generators and other group--theoretical
factors can be found in App.~\ref{app:a}. The explicit form of the
generators used in this work is given in App.~\ref{app:b}.  If the
link variables in the fundamental representation are written as:
\begin{equation}
U(x,\mu)=\exp\left[i \omega^a(x,\mu) T_F^a\right],
\end{equation}
then the link variables in the representation $R$ are given by:
\begin{equation}
U^R(x,\mu) = \exp\left[i \omega^a(x,\mu) T_R^a\right],
\end{equation}
where the functions $\omega^a(x,\mu)$ are {\em the same} in both
equations.  The explicit relation between link variables in different
representations can be worked out explicitly for each individual case
(see App.~\ref{app:b}).  For instance for the adjoint representation
of SU($N$) we have the well--known formula:
\begin{equation}
\label{eq:adj}
U^\mathrm{adj}_{a,b} = \frac12 \mathrm{tr}\left[T^a_F U T^b_F
  U^\dagger \right].
\end{equation}

The fermionic lattice action: 
\begin{equation}
S_f = a^4 \sum_x \bar\psi(x) D_m \psi(x)
\end{equation}
is quadratic in the fermionic fields and therefore its contribution to
the partition function can be computed exactly, yielding the fermionic
determinant, $\det D_m$. In numerical simulations, the latter
determinant is conveniently replaced by the determinant of the
Hermitean Dirac operator $Q_m$, defined as usual as:
\begin{equation}
\label{eq:Dirac5}
Q_m = \gamma_5 D_m.
\end{equation}
The fermionic determinant for an even number $n_f$ of degenerate
flavours can be represented by introducing complex pseudofermionic (bosonic)
fields $\phi,\phi^\dagger$:
\begin{equation}
\left(\det D_m\right)^{n_f} =
\left(\det Q_m\right)^{n_f} = 
\int \mathcal D \phi \mathcal D \phi^\dagger e^{-S_f},
\end{equation}
where the pseudofermionic action has the generic form:
\begin{equation}
  \label{eq:pseudof}
  S_f= a^4 \sum_x \phi^\dagger(x) \left(Q^2_m\right)^{-l} \phi(x),~~~~l=n_f/2.
\end{equation}
Note that we have used the same symbol $S_f$ to denote the fermionic
action written in term of the usual Grassmann variables and the
pseudofermionic action written in terms of bosonic fields that is used
in the numerical simulations; the meaning of the symbol will be
specified by the context in which it is used.  In the above expression
the square of $Q_m$ is used and not the matrix itself because $Q_m$ may
not be positive definite.  The sum over spin and color indices in the
above formulas has been omitted in order to simplify the notation.

For the gauge action we always use the Wilson action written in terms
of elementary $1\times 1$ plaquettes built of link variables {\em in
  the fundamental representation}:
\begin{equation}
  \label{eq:SGdef}
  S_g = \beta \sum_{\mu<\nu} \left( 1- \frac{1}{N} 
    \mathrm{Re\ tr} \mathcal{P}_{\mu\nu}\right) \, ,
\end{equation}
where $\beta=2N/g_0^2$ is the lattice bare coupling.  The algorithm
can be generalized to other gauge actions without any major
difficulty.

The partition function for the theory with $n_f$ flavors of dynamical
quarks can therefore be written as:
\begin{equation}
\label{eq:pathI}
Z=\int \mathcal D U \mathcal D \phi \mathcal D \phi^\dagger e^{-S_g(U)-S_f(U,\phi,\phi^\dagger)}.
\end{equation}
It is also possible to work exclusively with link
variables in the higher representation $R$, by writing the gauge
action as a function of the latter. We expect the two choices to yield
the same physics in the continuum limit, while a different phase
structure may appear at strong coupling. We will not investigate this
option here.

\section{Simulation algorithm}
\label{sec:alg}

Simulations for this work are performed using the RHMC
algorithm~\cite{Clark:2003na}.  This is an exact algorithm which is
flexible enough to cover all the cases of interest with good
performances, while still maintaining a relatively simple form.  The
algorithm itself is a variation of the HMC
algorithm~\cite{Duane:1987de} where rational approximations are used
to compute fractional powers of the fermionic matrix $Q_m^2$ appearing
for a generic number of flavors and pseudofermions.  Both the HMC and
RHMC set up a Markov chain of gauge field configurations yielding the
desired limiting distribution using the same steps.  The outline of
the algorithms can be summarised as follows: i) generate new
pseudofermion fields from a heat-bath; ii) evolve the gauge field
configuration following the flow of a fictitious Hamiltonian with
randomly chosen initial momenta for each link variable -- this step is
usually called molecular dynamic (MD) evolution; iii) perform a
Metropolis test at the of the MD trajectory to correct for errors in
the integration of the equation of motion at the previous step.

Our implementation of the HMC/RHMC is a straightforward adaptation
of the standard technique to the case under consideration. As usual the
bottleneck of the algorithm is the inversion of the Dirac operator
$Q_m^2$ that is needed during the MD evolution.  To reduce the number
of steps in the integration of the equation of motion and thus the
number of inversions, we use the second order Omelyan
integrator~\cite{Takaishi:2005tz} for the MD evolution with different
time steps for the gauge and fermion action.  Multiple pseudofermions
and the even-odd preconditioning of the fermionic determinant are also
used in this paper, but no other acceleration techniques are necessary
for the purpose of this work. More extensive simulations will require
a more careful study of these issues. 

In the rest of this section we will describe the modifications required to handle fermions 
in a generic representation and the implementation details used in our code.

\subsection{HMC molecular dynamics}
\label{sec:hmc}
In the following the expression for the force needed in the Molecular
Dynamics is generalized to the case of fermions in a generic
representation. Since simulations in arbitrary representations are
still at their early stages, we decide to derive the forces in detail,
in order to define the quantities that appear in the simulations as
clearly as possible. Introducing for each link variable a conjugate
momentum in the algebra of the gauge group:
\begin{equation}
\pi(x,\mu)=i \pi^a(x,\mu) T_F^a,
\end{equation}
a fictitious Hamiltonian $H$ is written on the group manifold as:
\begin{equation}
  \label{eq:hamilt}
  H=H_\pi+H_g+H_f \, ,
\end{equation}
where, assuming $n_f=2$:
\begin{eqnarray}
  H_\pi &=& \frac{1}{2} \sum_{x,\mu} \bigg( \pi(x,\mu) ,
  \pi(x,\mu) \bigg) = \frac{1}{2} T_F \sum_{a,x,\mu} \pi^a(x,\mu)^2 \, ,\\
  H_g &=& S_g = \beta \sum_{\mu<\nu} \left( 1- \frac{1}{N} 
    \mathrm{Re\ tr} \mathcal{P}_{\mu\nu}\right) \, ,\\
  H_f &=& S_f = \sum_x \phi^\dagger(x) \left[ Q_m^2 - s \right]^{-1}
  \phi(x) \, \label{eq:HFHMC}.
\end{eqnarray}
Since the momenta are conjugate to the link variables, they are
defined in the fundamental representation. 

The cases where $n_f\neq 2$ are dealt using the RHMC algorithm
described in the next section. Also note that an arbitrary shift
$s$ has been included in the fermionic action which will be useful
for the discussion of the RHMC algorithm.

Denoting by $\tau$ the fictitious time of this Hamiltonian
system, the equation of motion are given by:
\begin{eqnarray}
  \label{eq:eom1}
  \frac{d}{d\tau}U(x,\mu)&=&\pi(x,\mu) U(x,\mu) \label{eq:Uevol}\\
  \label{eq:eom2}
  \frac{d}{d\tau}\pi(x,\mu)
	&=& F(x,\mu), 
\end{eqnarray}
where the RHS of the second equation above defines the force
$F(x,\mu)$ that drives the time evolution of the momenta.  

For a generic infinitesimal variation of a link variable:
\begin{equation}
\label{eq:deltaU}
\delta U(x,\mu) = \delta\omega(x,\mu) U(x,\mu),
\end{equation}
where $\delta\omega(x,\mu)=i\delta\omega^a(x,\mu) T_F^a$ is an element
of the algebra, $F(x,\mu)$ is obtained from the
corresponding variation of the action through the equations:
\begin{eqnarray}
F(x,\mu) &=& F_g(x,\mu) + F_f(x,\mu) , \nonumber \\
\delta S_g &=& -\left(\delta\omega,F_g\right), \nonumber \\
\delta S_f &=& -\left(\delta\omega,F_f\right). \nonumber
\end{eqnarray}
Since we use link variables in the fundamental representation for the
gauge action $S_g$, the variation of the latter is the usual one that
appears in the HMC molecular dynamics evolution:
\begin{eqnarray}
  \delta S_g &=& - \frac{\beta}{N} \sum_{x,\mu,a}  \,
  \delta\omega^a(x,\mu)\, \mathrm{Re\ tr} \left[ i T^a_F U(x,\mu)
    V^\dagger(x,\mu) \right] \, , \label{eq:forceG} 
\end{eqnarray}
where $V(x,\mu)$ is the sum of the forward and backward staples around
the link $U(x,\mu)$.

The fermionic force is obtained from the variation
of the fermionic action as follows. Starting from:
\begin{eqnarray}
\delta S_f = -\ \phi^\dagger (Q_m^2 - s)^{-1}
\delta (Q_m^2) (Q_m^2 - s)^{-1} \phi \, , \label{eq:FF1}
\end{eqnarray}
let us define:
\begin{eqnarray}
\eta &=& (Q_m^2 - s)^{-1} \phi \, , \label{eq:etahmc}\\
\xi &=& Q_m \eta \, ;
\end{eqnarray}
using the fact that the matrix $(Q_m^2-s)$ is hermitean, we can
rewrite Eq.~(\ref{eq:FF1}) as
\begin{equation}
\delta S_f = - 2 \mathrm{Re}\left[ \xi^\dagger \delta (Q_m) \eta \right] \, . \label{eq:FF2}
\end{equation}
Inserting the explicit form of $Q_m$, Eq.~(\ref{eq:Dirac5}), into
Eq.~(\ref{eq:FF2}) we obtain
\begin{multline}
\label{eq:deltaSF}
\delta S_f =  \mathrm{Re\ }\sum_{x,\mu} \left[ 
\xi(x)^\dagger \delta U^R(x,\mu) \gamma_5 (1-\gamma_\mu) \eta(x+\mu)+
\right. \\ 
\left. + \eta(x)^\dagger \delta U^R(x,\mu) \gamma_5 (1-\gamma_\mu) \xi(x+\mu)\right]
\end{multline}
where we are implicitly summing over spin and color indices.
We can now write the variation of $U^R$ as:
\begin{equation}
\label{eq:deltaUR}
\delta U^R (x,\mu) = \delta\omega^R(x,\mu) U^R(x,\mu) = i \,
\delta\omega^a(x,\mu) T^a_R U^R(x,\mu), 
\end{equation}
where the $\delta\omega^a(x,\mu)$ in the above equation are the same
functions that define the variation of the gauge link in
Eq.~(\ref{eq:deltaU}). Eq.~(\ref{eq:deltaSF}) and~(\ref{eq:deltaUR})
yield:
\begin{multline}
\delta S_f = \sum_{x,\mu,a}  
\delta\omega^a(x,\mu)\, \mathrm{Re\ tr}_{c,s} \left[ iT^a_R U^R(x,\mu) \gamma_5 \right. \times\\
\times\left. (1-\gamma_\mu) \left\{ \eta(x+\mu)\otimes\xi(x)^\dagger + 
    \xi(x+\mu)\otimes\eta(x)^\dagger \right\} \right] \, . \label{eq:forceF}
\end{multline}
Here the symbol $\mathrm{tr}_{c,s}$ indicates the trace over color
and spin.

For the sake of convenience we introduce the following projectors
$P^a_R$ over the algebra in the representation $R$:
\begin{equation}
P^a_R ( F ) = - \frac{1}{T_F} \mathrm{Re\ tr}_c \left[ i T^a_R F \right] \, ,
\end{equation}
and the following trace operator:
\begin{multline}
\mathrm{Tr}_{x,\mu} (\eta,\xi) = \mathrm{tr}_s \left[\gamma_5 (1-\gamma_\mu)  \left\{ \eta(x+\mu)\otimes\xi(x)^\dagger +\right.\right.\\
\left.\left. +\xi(x+\mu)\otimes\eta(x)^\dagger \right\} \right] \, .
\end{multline}
The forces are then given by:
\begin{align}
F^a_G(x,\mu) &= - \frac{\beta}{N} P^a_F \left( U(x,\mu) V^\dagger(x,\mu) \right) \, ,\\
F^a_F(x,\mu) &= P^a_R \left( U^R(x,\mu) \mathrm{Tr}_{x,\mu} (\eta,\xi) \right)\, .\label{HFFORCE}
\end{align}

Note that when $R$ is chosen to be the fundamental representation, the
usual expressions for the fermionic force are recovered.
Explicit expressions for the forces had already
been computed for the case of fermions in the adjoint representation
of the gauge group SU(2) in Ref.~\cite{Donini:1996nr}. As an analytic
check, we have verified that our result above agrees with Eq.~(16) in
Ref.~\cite{Donini:1996nr}.

\subsection{RHMC implementation}
\label{sec:rhmc}

The fermionic part of the HMC hamiltonian, for $n_f$ degenerate 
fermions and $N_{pf}$ pseudo\-fermions, can be written as:
\begin{equation}
H_f = \sum_{k=1}^{N_{pf}} \phi_k^\dagger ( Q_m^2 )^{-l_k}
\phi_k \,\, ;\,\, \sum_{k=1}^{N_{pf}} l_k = \frac{n_f}{2}\, . \label{HFN}
\end{equation}
For the sake of simplicity we will set all the $l_k$ to be equal:
\begin{equation}
\forall k,\,\, l_k = \frac{n_f}{2N_{pf}}\, .
\end{equation}
The above decomposition is used in the RHMC algorithm~\cite{Clark:2003na}, where rational
approximations are used to compute the fractional powers of the
positive definite fermion matrix $Q_m^2$. Even
though we will work at fixed $n_f=2$ in this paper, the RHMC is
particularly useful if one wants an algorithm that can be easily
generalized to an arbitrary number of fermions. Three different
rational approximations are used for this implementation.

The first rational approximation is required in the heat--bath update of
pseudofermions.  In order to generate pseudofermions distributed as in
Eq.~(\ref{HFN}), a simple two-step process is used.  For each
pseudofermion we first generate a gaussian distributed field
$\tilde\phi_k$ with probability:
\begin{equation}
P(\tilde\phi_k)\propto \exp [ -\tilde\phi_k^\dagger \tilde\phi_k ] \, ,
\end{equation}
and then we set:
\begin{equation}
\phi_k = (Q_m^2)^{\frac{l_k}{2}} \tilde\phi_k \, ,
\end{equation}
making use of the fact that $(Q_m^2)$ is hermitean (notice the plus
sign in the exponent.)  The RHMC algorithm uses a rational
approximation to compute the above quantities (we need only one
approximation since all $l_k$ are equal):
\begin{eqnarray}
 r_{a}(Q_m^2) &=& \alpha_0^a +
\sum_{n=1}^{d_{1}} \alpha_n^a ( Q^2_m - s_n^a )^{-1} \simeq ( Q_m^2 )^{\frac{l_k}{2}} \, .
\end{eqnarray}

The second rational approximation is used to approximate Eq.~(\ref{HFN}) 
during the molecular dynamics evolution (as before only one
approximation is needed because all $l_k$ are equal):
\begin{eqnarray}
H_F &=& \sum_{k=1}^{N_{pf}} \phi_k^\dagger r_{b}( Q_m^2
)\phi_k \, , \label{HFRHMC}\\ 
r_{b}(Q_m^2) &=& \alpha_0^b + \sum_{n=1}^{d_{2}} \alpha_n^b ( Q^2_m -
s_n^b )^{-1} \simeq ( Q_m^2 )^{-l_k}  \, .
\end{eqnarray}
Using the formulas derived in Section~\ref{sec:hmc}, it is easy to write
the force corresponding to Eq.~(\ref{HFRHMC}).  In fact,
Eq.~(\ref{HFRHMC}) is nothing but a sum of terms of the form
Eq.~(\ref{eq:HFHMC}) once we put $l=1$, $s=s_n^b$.  The RHMC
force will be then given by a sum over $n=1,\dots,d_2$ of terms given
by Eq.~(\ref{HFFORCE}) multiplied by a factor $\alpha_n^b$.  
It is possible to compute all the $\eta$'s defined in Eq.~(\ref{eq:etahmc}) 
corresponding to different $s_n^b$ simultaneously
with a multi-shift inverter.

The third rational approximation is used for the Metropolis test.
Starting from Eq.~(\ref{HFN}), for each pseudofermion we can rewrite:
\begin{equation}
\phi_k^\dagger ( Q_m^2 )^{-l_k}\phi_k = \left\|
(Q_m^2)^{-\frac{l_k}{2}} \phi_k \right\|^2\, ,
\end{equation}
where we used the property that $Q_m^2$ is hermitean.
The rational approximation needed in this case is:
\begin{eqnarray}
r_{c}(Q_m^2) &=& \alpha_0^c +
\sum_{n=1}^{d_{3}} \alpha_n^c ( Q^2_m - s_n^c )^{-1} \simeq ( Q_m^2 )^{-\frac{l_k}{2}} \, .
\end{eqnarray}
Notice that if $d_1=d_3$ the coefficients for the two approximations
$r_a$ and $r_c$ can each be obtained from the other.

The coefficients $\alpha_n$, $s_n$ appearing
in the rational approximations are computed using the Remez algorithm. 
In this
implementation we do not compute the coefficients ``on the fly'' but instead
the required values are taken from a look-up table that has been
precomputed, according to the following prescription.

First note that we need to compute rational approximations for a
function $f(x)$ of the form $f(x)=x^l$ and the approximation must be
accurate over the spectral range of the operator $Q_m^2$.  To simplify
the computation of the table we note that the following proposition
holds: if $f(x)$ is a homogeneous function of degree $l$ and $r(x)$ is
an optimal (in the sense of relative error) rational approximation to
$f(x)$ over the interval $[\epsilon,\mathrm{h}]$ to a given accuracy
then $r(kx)/k^l$ is an optimal rational approximation for the same
function and the same accuracy over the interval
$[\epsilon/k,\mathrm{h}/k]$. Moreover the coefficients of this
``rescaled'' rational approximation are easily obtained from that of
the original approximation.  A simple corollary is that
we can divide the optimal rational 
approximations of a given homogeneous function $f(x)$
with the same relative precision, in classes labeled by the ratio
$\epsilon/\mathrm{h}$. Within each of these classes the coefficients 
of the rational approximations are easily related to each other, so that we
only need to compute one rational approximation in each class. 
Thus all that is needed in practice is a table containing the coefficients 
of rational approximations for each of these classes, for
each function $f(x)$ which we want to approximate and for each relative precision
which is required. 
At run-time this table is used to generate optimal rational
approximations rescaling the precomputed coefficients to the desired
interval containing the spectrum of the matrix $Q_m^2$.  This interval
is obtained by computing the maximum and minimum eigenvalue of $Q_m^2$
on each configuration when needed. 

\subsection{Even-Odd preconditioning}
As already pointed out above, the time required for the inversions of
the Dirac operator is one of the dominant contributions to the total
cost of the simulation.  The convergence of such inversions can be
improved using an appropriate preconditioning.  The idea behind
preconditioning is to rewrite the fermionic determinant as a
determinant (or product of determinants) of one (or more) better
conditioned matrix (matrices) than the original Dirac operator. Very
effective preconditionings are at the heart of the recent progress in
simulations of QCD with dynamical fermions. For the scope of this
work, we use a simple \textit{even--odd}
preconditioning~\cite{DeGrand:1990dk}. We review the main features of
even--odd preconditioning here in order to explain the modifications
that are required when considering higher--dimensional
representations. We divide the lattice in a sublattice of even points
$\Lambda_e$ and another sublattice of odd points $\Lambda_o$, and we
rewrite the Dirac operator $D_m$ as a block matrix:
\begin{equation}
D_m = \begin{pmatrix}
4+m& D_{eo}\\
D_{oe} &4+m
\end{pmatrix}\,\,\, ,
\end{equation}
where each block has a dimension half that of the original Dirac
matrix. The diagonal blocks connecting sites with the same parity are
proportional to the identity matrix, while off-diagonal blocks connect
sites with opposite parity. We have (since $D_m$ is
$\gamma_5$-hermitean):
\begin{equation}
\gamma_5 D_{eo} \gamma_5 = D_{oe}^\dagger\,\, . 
\end{equation}
The determinant of the Dirac matrix $D_m$ can be rewritten as:
\begin{equation}
{\rm det\ } D_m = {\rm det} \left[ (4+m)^2 - D_{eo} D_{oe} \right] \equiv {\rm det\ } D^{eo}_m\,\, ,
\end{equation}
using the well known expression for the determinant of a block matrix.
Since the determinant of $D_m$ and of $D_m^{eo}$ are the same the
latter can be used in numerical simulations. Note that the even-odd
preconditioned matrix only connects sites with the same parity thus it
has only half of the size of the original Dirac matrix and, like
$D_m$, it is $\gamma_5$-hermitean. We define as before the hermitean
matrix $Q_m^{eo}\equiv \gamma_5 D_m^{eo}$, which will be used in
practice.

The formulation of the HMC algorithm does not change and the only
difference is that pseudofermions fields are now only defined on half
of the lattice sites, conventionally the even sites in what follows.
We now derive the explicit expression for the fermionic force for the
preconditioned system described by the hamiltonian:
\begin{eqnarray}
H_f &=& \phi_e^\dagger \left[ (Q^{eo}_m)^2 - s \right]^{-1} \phi_e \,\, ,
\end{eqnarray}
where as before we are assuming $n_f=2$ or a rational approximation of
the actual fractional power function; the suffix in $\phi_e$ is an
explicit reminder that the pseudofermion field is only defined on even
sites.  Eq.~(\ref{eq:FF2}) is unchanged :
\begin{equation}
\delta S_f = - 2 \mathrm{Re}\left[ \xi_e^\dagger \delta (Q^{eo}_m)
  \eta_e \right] \, ,
\label{eq:FFPRE}
\end{equation}
where as before we have defined:
\begin{eqnarray}
\eta_e &=& ((Q^{eo}_m)^2 - s)^{-1} \phi_e \, , \\
\xi_e &=& Q^{eo}_m \eta_e \, .
\end{eqnarray}
The explicit form of $Q_m^{eo}$ must be used at this point. We have:
\begin{equation}
\delta (Q^{eo}_m) = -\gamma_5 (\delta D_{eo} D_{oe} + D_{eo}\delta
D_{oe} )\,\, .
\label{eq:QPREDOT}
\end{equation}
Defining
\begin{eqnarray}
\eta_o &=& D_{oe} \eta_e \, , \\
\xi_o &=& D_{oe} \xi_e \, ,
\end{eqnarray}
and inserting Eq.~(\ref{eq:QPREDOT}) into Eq.~(\ref{eq:FFPRE}), we arrive at the same expression as before for the variation of the action, but with a minus sign:
\begin{eqnarray}
\delta{S_F} = &-& \sum_{\mu,x} \mathrm{Re\ tr}_c \left[\delta U^R(x,\mu) \mathrm{Tr}_{x,\mu}(\eta,\xi) \right]
\label{eq:FORPRE}
\end{eqnarray}
From Eq.~(\ref{eq:FORPRE}) and proceeding as before we arrive at the final expression for the force. This again coincides with Eq.~(\ref{HFFORCE}) but with the opposite sign:
\begin{equation}
F_F^a(x,\mu) = - P^a_R \left( U^R(x,\mu) \mathrm{Tr}_{x,\mu}(\eta,\xi) \right)
\end{equation}

\subsection{Some details on the implementation}
As the algorithm is implemented for the first time, we describe in
this section some general features of our code and the technical
solutions adopted to manage fermions fields defined in an arbitrary
representation. 

At this early stage in our physics studies we prioritize the
flexibility of the code over the optimization aspects. The code is
written in ANSI C for maximum portability to different architectures.
This choice is justified, in our opinion, by the fact that, on this
relatively small scale, C codes maintain a much higher overall
simplicity and clarity compared to e.g. C++ codes with a comparable
low overhead.

The major difference with respect to conventional codes -- i.e. codes
for simulations of the SU(3) gauge group with fundamental fermions --
is obviously that all operations involving gauge links or
pseudofermions fields must be able to handle matrices and vectors of
arbitrary dimension.  In the generic case of four--dimensional SU($N$)
with fermions in a representation $R$, link matrices for the gauge
action have dimension $N\times N$, the represented gauge links have
dimension $d_R\times d_R$ and fermion fields have dimension $4
d_R$. These numbers indicate how the cost of the simulation scales
both in memory and in the number of floating point operations per
application of the Dirac operator as $N$ and $R$ are varied. One
simple way to handle this objects would be to have functions that
implement operations on them taking as an additional parameter the
dimensions of the objects themselves.  However we found it more
flexible to use the following approach. We use preprocessor macros
instead of functions\footnote{Inline functions could work just as
well, however the handling of inlining is a feature which depends on
the compiler.}. We fix the gauge group and the representation $R$ at
compile time as this gives the opportunity to more optimizations
(e.g. partial unrolling of loops)\footnote{This also means that
recompiling is needed whenever the number of colors or the fermions
representation is changed. However since the total compilation time,
including the automatic generations of required header, is less than
one minute and independent on $N$, this can hardly be a problem.}. All
of the required macros are automatically generated in a pre-compiling
step. We use a simple Perl program for this purpose whose input
parameters are the number of colors $N$ and the representation $R$ and
whose output is a C header file containing the macro functions. The
whole process of header generation and compilation is managed through
a custom Makefile system and it is thus completely transparent to the
end user.

A second difference with respect to conventional codes, is that it is
convenient, although not strictly necessary, to have a second copy of
the gauge field in the representation $R$. This is only used in the
computation of the fermionic force, see Eq.~(\ref{HFFORCE}), and it is
also useful to impose boundary conditions on the fermion fields.
Every time the gauge field is updated during the molecular dynamics
evolution, the represented gauge field needs to be recomputed.

At this early stage of simulations, we chose to use double precision 
everywhere in the code to reduce the risk of problems arising from numerical accuracy.
This is done at the expense of optimal performance, which we feel is justified in this study of the algorithm implementation and behavior. 
As the source of random numbers, we use the RANLUX generator~\cite{Luscher:1993dy}.

\section{Behavior of the algorithm}
\label{sec:res}
In this section we shall discuss some simple tests of the RHMC
algorithm with the aim to show the correctness of our implementation.
The majority of the tests performed for this purpose were run on the lattice
T2-B11 (see Tab.~\ref{tab:T2runs} below) which corresponds to our
biggest volume and the lightest mass at $\beta=2$. 
For all the tests in
this section we have used even-odd preconditioning and two
pseudofermion fields.

\subsection{Preliminary Tests}

A simple test of the algorithm is to look at the expectation value
$\langle \exp(-\Delta H)\rangle$, where $\Delta H$ is the difference
of the values of the Hamiltonian $H$ at the beginning and at the end
of a trajectory in the Molecular Dynamics evolution. This expectation
value should always be 1 (a result known as Creutz
equality~\cite{Creutz:1988wv}).  Moreover since the
RHMC is an exact algorithm, i.e. there are no corrections due to the
errors occurring in the Molecular Dynamics integration, a good test of
the algorithm is to check whether or not there is any dependence of
any quantity on the integration step size $\Delta\tau$.  The average
plaquette $\langle P\rangle$ is a good candidate to this purpose
because it can be measured with high accuracy.

We show in Fig.~\ref{fig:dtdep} the quantities $\langle \exp(-\Delta
H)\rangle$ (upper panel) and the average plaquette (lower panel) for
four different step sizes of the MD integration.  No statistically
significant deviations from the expected behavior are seen.  The value
of $\langle P\rangle$ is consistently independent on $\Delta\tau$ and
the Creutz's equality $\langle \exp(-\Delta H)\rangle=1$ is
satisfied. Notice also that for the latter quantity the error bars
become smaller as $\Delta\tau$ goes to zero, as expected from the fact
that $\langle\Delta H\rangle$ is also vanishing in the same limit.
%
\FIGURE{
  \centering
  \epsfig{file=fig1.eps,width=\columnwidth,clip}  
  \caption{Dependence of $\langle \exp\left(\Delta H\right)\rangle$ and $\langle P \rangle$ on the time--step used for the MD integration.}
  \label{fig:dtdep}
}
%
%
\FIGURE{
  \centering
  \epsfig{file=fig2.eps,width=\columnwidth,clip}  
  \caption{The expectation value $\langle \Delta H\rangle$ is
    proportional to $(\Delta\tau)^4$ (upper panel) consistently with
    the use of a second order integrator (the red curve shows the one
    parameter fit). The acceptance probability $P_{acc}$ as measured
    in the test runs (lower panel) is correctly described by the
    expected large volume behavior
    $P_{acc}=\mathrm{erfc}(\sqrt{\langle\Delta H\rangle}/2)$ (solid
    curve, not a fit).}\label{fig:dHacc}
}
This is explicitly shown in the upper panel of
Fig.~\ref{fig:dHacc}. As expected for a second order integrator (see
\cite{Takaishi:1999bi}), the quantity $\langle\Delta H\rangle$ is
proportional to $\Delta\tau^4$ with very good
accuracy (the red curve in the plot is the best
fit to that functional form with one free parameter).  
The acceptance probability $P_{acc}$ can also be used a
test of the correctness of the algorithm.  In fact in the large volume
limit this probability is given by
$P_{acc}=\mathrm{erfc}(\sqrt{\langle\Delta
  H\rangle}/2)$~\cite{Gupta:1990ka}.  We show the
measured value of the acceptance rate as a function of the average
Hamiltonian variation in the lower panel of Fig.~\ref{fig:dHacc}. The
same figure shows also the predicted behavior (solid curve).  We found
a convincing agreement with the expectations.

As a last test in this section we measured reversibility violations in
the MD. We remind the reader that this is a necessary condition for
the correctness of the algorithm.  The discussion will follow that of
Ref.~\cite{Joo:2000dh}.  At the ``microscopic'' level,
reversibility violations can occur when updating the gauge variables
and in the momenta update during the MD evolution.  We deal with the
first source of non-reversibility using the trick suggested in
Ref.~\cite{Luscher:2005rx}: the link update Eq.~(\ref{eq:Uevol})
is implemented by left multiplication of an exactly unitary matrix
$E[\pi(x,\mu)]$ such that $E[\pi(x,\mu)] E[-\pi(x,\mu)] = 1$ and which is
also an approximation to the exponential map. In this way no possible
reversibility violations can arise (within machine precision) in the
link update.  For the momenta update, local reversibility is guaranteed
by the use of double precision and the requirement of a small relative
residue ($10^{-7}$ in our simulations) for the force calculation. The
solution of the linear system required for the force calculation
itself is always started from the same trial solution.  At the
``global'' level we measured the reversibility violations through the
quantity $|\delta H|$ defined as the average difference between
the Hamiltonian of the starting configuration and the one obtained
evolving the system forward for a unit of Monte Carlo time and then
back (flipping the momenta at $\tau=1$) to the original position in
phase space. This measure of reversibility violations is shown in
Fig.~\ref{fig:revtest}. The global violations to reversibility appear
to be very small and
independent on $\Delta\tau$ and hence also on $\langle \Delta
H\rangle$.
%
\FIGURE{
  \centering
  \epsfig{file=fig3.eps,width=\columnwidth,clip}  
  \caption{Reversibility test for several values of the time--step
    used for the MD integration.}
  \label{fig:revtest}
}

\subsection{Checks for SU(3) with $n_f=2$ in the fundamental representation}
The routines that perform linear algebra operations in color
space for arbitrary representations are generated automatically by
a Perl program. As a first non--trivial test of our implementation, our code
should correctly reproduce the results for simulations in the fundamental
representation of SU(3). The code has been benchmarked for SU(3) with
$n_f=2$ against simulations obtained using the DD--HMC
algorithm~\cite{Luscher:2005rx} on small
lattices. Fig.~\ref{fig:cfrDDHMC} shows the thermalization of the
plaquette at $\beta=6.0$. On small lattices, the DD--HMC only updates
a small fraction of the links in the system, and therefore the
thermalization is much slower in units of MC trajectories. 
To compare the two time--histories we
have therefore rescaled the trajectory number for the DD--HMC data.
The figure shows a very good  agreement between the two simulations, 
in particular for the equilibrium value of the plaquette.
%
\FIGURE{
  \centering
  \epsfig{file=fig4.eps,width=\columnwidth,clip}
  \label{fig:cfrDDHMC}
  \caption{Time--history for the thermalization of the plaquette in
    the SU(3) theory with two flavours in the fundamental
    representation on a $16^4$ lattice at $\beta=5.6$ and
    $\kappa=0.15750$. The black line shows the evolution of the
    plaquette value using our HMC algorithm, while the red line
    represents the same quantity using the DD-HMC
    algorithm. Trajectory number has been scaled by a factor of 7 for
    the DD-HMC data.}
}

Another test of the algorithm is obtained by simulating the SU(3)
theory with two flavors in the fundamental representation and in the
two--index antisymmetric one. For the theory with three colors the two
representations are equivalent and related by charge conjugation. We
have compared the outcomes of two simulations on $4^3 \times 8$ lattice
at $\beta=5.6$ and $\kappa=0.15600$. The time--history of the
plaquette and its probability distribution for the two simulations are
summarized in Fig.~\ref{fig:cfrfund2AS}. Once again we find a very
convincing agreement between the two simulations.

\FIGURE{
  \centering
  \epsfig{file=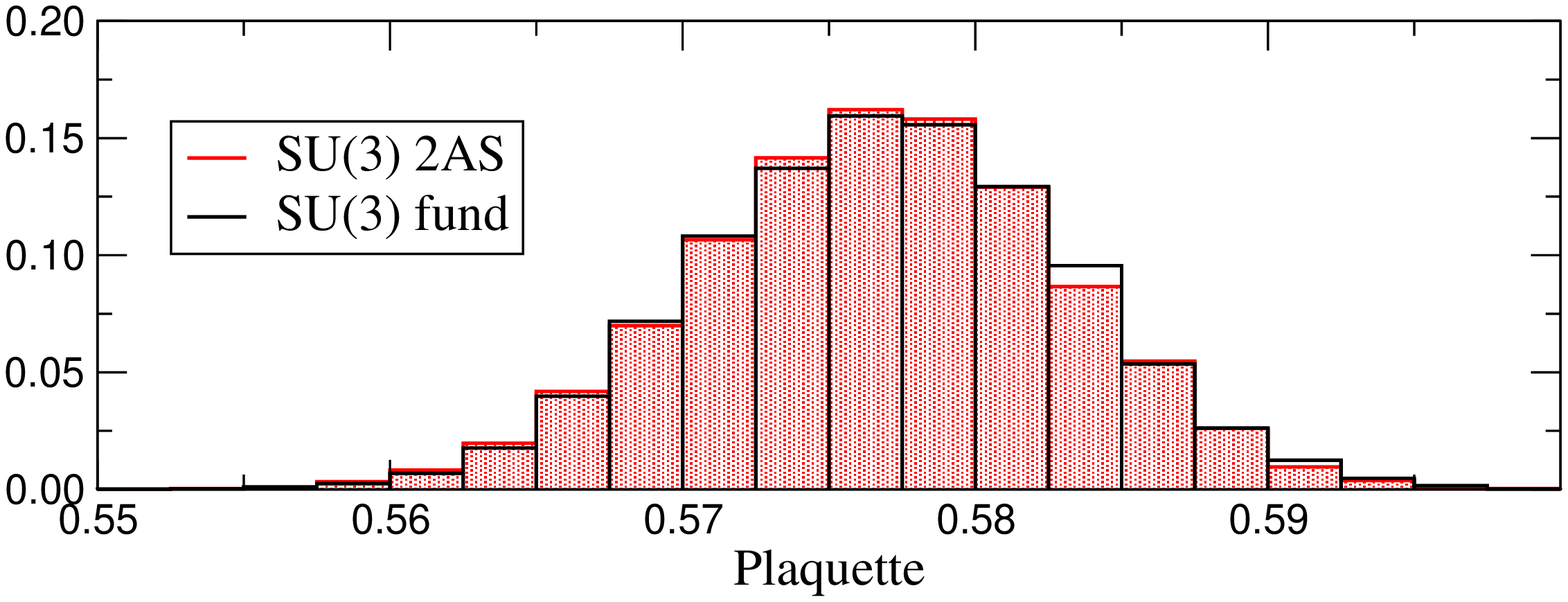,width=\columnwidth,clip}  
  \caption{Probability distribution of the plaquette. 
  The black (respectively red) curve refers to data
  from a simulation of the SU(3) gauge theory on a $4^3 \times 8$
  lattice at $\beta=5.6$, $\kappa=0.15600$ with two fermions in the
  fundamental representation (respectively the two--index
  antisymmetric).}
  \label{fig:cfrfund2AS}
}

\subsection{SU(2) $n_f=2$ two--index symmetric}
The theory with two colors and two Dirac fermions in the two--index
symmetric ($2S$) representation has been proposed as a
phenomenologically relevant candidate for a minimal walking
technicolor theory. For the specific case of the SU(2) color group,
the $2S$ representation is equivalent to the adjoint representation
and is therefore real, which makes it less intensive from the
computational point of view. Analytical results for the $\Lambda$
parameter, the renormalization constant, and the possibility of an
Aoki phase at finite lattice spacing have been presented in
Ref.~\cite{DelDebbio:2008wb}, where the theory was denoted T2. We
will adopt the same conventions here. Preliminary results from lattice
simulations of this theory were presented in
Ref.~\cite{Catterall:2007yx}. In this subsection we build upon this
previous experience and provide some more extensive description of the
behaviour of numerical simulations in the space of bare
parameters. The list of runs is summarized in Tab.~\ref{tab:T2runs} for inverse coupling $\beta=2.0$ and in Tab.~\ref{tab:T3runs} for inverse coupling $\beta=2.25$.
%
\TABLE{
\centering
\begin{tabular}{ccccccc}
\hline
lattice & $V$ & $\kappa$ & $-am_0$  & $N_\mathrm{traj}$ &  $\langle P\rangle$ &
$\tau$ \\
\hline
T2-A1 & $8\times 4^3$ & 0.12500 & 0.0  & 28800 & 0.5093(14)  & 2.9(0.4) \\
T2-A2 & $8\times 4^3$ & 0.14286 & 0.5  & 28800 & 0.5163(16)  & 3.1(0.5) \\
T2-A3 & $8\times 4^3$ & 0.15385 & 0.75 & 28800 & 0.5235(18)  & 3.1(0.5) \\
T2-A4 & $8\times 4^3$ & 0.16667 & 1.0  & 28800 & 0.5373(20)  & 6.0(1.2) \\
T2-A5 & $8\times 4^3$ & 0.18182 & 1.25 & 27200 & 0.5742(37) & 12.0(3.6) \\
T2-A6 & $8\times 4^3$ & 0.18382 & 1.28 & 25600 & 0.5850(50) & 22.3(9.3) \\
T2-A7 & $8\times 4^3$ & 0.18587 & 1.31 & 41600 & 0.6013(55) & 48.3(23.3)\\ 
T2-A8 & $8\times 4^3$ & 0.18657 & 1.32 & 51200 & 0.6159(58) & 40.7(16.3)\\ 
\hline
T2-A1$^\prime$ & $8\times 4^3$ & 0.12500 & 0.0 & 3000  & 0.5094(45)  & 2.7(1.2) \\
\hline
T2-B7  & $16\times 8^3$ & 0.18587 & 1.31  & 3200 & 0.5951(42) & 5.8(3.6) \\
T2-B8  & $16\times 8^3$ & 0.18657 & 1.32  & 1600 & 0.6040(56) & 9.0(9.6) \\
T2-B9  & $16\times 8^3$ & 0.18692 & 1.325 & 2240 & 0.6107(53) & 4.2(2.6) \\
T2-B10 & $16\times 8^3$ & 0.18727 & 1.33  & 1100 & 0.6168(73) & 2.6(1.8) \\
T2-B11 & $16\times 8^3$ & 0.18797 & 1.34  & 3840 & 0.6347(58) & 13.6(11.5) \\
\hline
\end{tabular}
\label{tab:T2runs}
\caption{List of runs for the $T2$ theory at inverse coupling $\beta=2.0$. 
  The standard runs are
  performed using link variables in the real
  adjoint representation. The primed run is performed with link
  variables in the $2S$ representation that are constructed in an
  algorithmic way as complex $3 \times 3$ matrices. In order to
  simplify the comparison with the runs in
  Ref.~\cite{Catterall:2007yx} we list both the values of $\kappa$ and
  $m$}
}
\WTABLE{
\centering
\begin{tabular}{cccccccc}
\hline
lattice & V & $-am_0$ & $N_{traj}$ & $\langle P\rangle$ & $\tau$ & $\lambda$ & $\tau_\lambda$ \\
\hline
T2-C0 & $16\times 8^3$ & 0.95 & 7601 & 0.63577(16) & 5.45(72) & 3.582(13) & 8.6(1.4)  \\
T2-C1 & $16\times 8^3$ & 0.975 & 7701 & 0.63843(15) & 5.43(71) & 2.982(12) & 6.65(96)  \\
T2-C2 & $16\times 8^3$ & 1 & 7801 & 0.64136(15) & 5.10(64) & 2.427(11) & 6.28(88)  \\
T2-C3 & $16\times 8^3$ & 1.025 & 7801 & 0.64463(15) & 4.29(50) & 1.894(10) & 6.07(84)  \\
T2-C4 & $16\times 8^3$ & 1.05 & 7801 & 0.64793(15) & 3.48(36) & 1.4596(79) & 4.39(52)  \\
T2-C5 & $16\times 8^3$ & 1.075 & 6400 & 0.65179(16) & 2.99(32) & 1.0692(74) & 4.27(55)  \\
T2-C6 & $16\times 8^3$ & 1.1 & 6400 & 0.65566(16) & 3.28(37) & 0.7564(60) & 3.77(45)  \\
T2-C7 & $16\times 8^3$ & 1.125 & 7073 & 0.66037(15) & 2.99(30) & 0.4854(43) & 3.03(31)  \\
T2-C8 & $16\times 8^3$ & 1.15 & 6400 & 0.66550(16) & 3.31(37) & 0.2779(31) & 2.80(29)  \\
T2-C9 & $16\times 8^3$ & 1.175 & 6400 & 0.67177(17) & 3.24(36) & 0.1351(18) & 2.80(29)  \\
\hline
T2-D0 & $24\times 12^3$ & 0.95 & 10201 & 0.635310(59) & 6.16(74) & 3.5058(50) & 3.08(26)  \\
T2-D1 & $24\times 12^3$ & 1 & 8652 & 0.640998(64) & 4.92(58) & 2.4218(44) & 3.10(29)  \\
T2-D2 & $24\times 12^3$ & 1.05 & 7819 & 0.647633(70) & 6.79(99) & 1.4936(51) & 5.80(78)  \\
T2-D3 & $24\times 12^3$ & 1.075 & 7186 & 0.651630(68) & 4.61(58) & 1.0553(40) & 4.95(64)  \\
T2-D4 & $24\times 12^3$ & 1.1 & 6393 & 0.655827(76) & 4.09(51) & 0.7202(30) & 7.8(1.3)  \\
T2-D5 & $24\times 12^3$ & 1.125 & 6200 & 0.660588(75) & 3.97(50) & 0.4419(22) & 5.98(91)  \\
T2-D6 & $24\times 12^3$ & 1.15 & 1599 & 0.66588(15) & 3.71(90) & 0.2271(31) & 6.6(2.1)  \\
T2-D7 & $24\times 12^3$ & 1.175 & 5582 & 0.672074(79) & 4.22(58) & 0.08641(90) & 3.78(49)  \\
T2-D8 & $24\times 12^3$ & 1.18 & 4081 & 0.673474(92) & 4.01(63) & 0.06561(92) & 10(2.5)  \\
T2-D9 & $24\times 12^3$ & 1.185 & 4201 & 0.675094(93) & 3.42(49) & 0.05196(71) & 3.53(51)  \\
T2-D10 & $24\times 12^3$ & 1.19 & 3501 & 0.67663(10) & 4.15(70) & 0.03985(61) & 5.2(1.0)  \\
\hline
T2-E0 & $32\times 16^3$ & 1.15 & 5446 & 0.665894(44) & 3.32(40) & 0.2227(10) & 3.05(36)  \\
T2-E1 & $32\times 16^3$ & 1.175 & 2192 & 0.672235(73) & 2.80(50) & 0.07036(90) & 5.9(1.5)  \\
T2-E2 & $32\times 16^3$ & 1.18 & 4606 & 0.673657(49) & 3.46(47) & 0.05167(50) & 6.1(1.1)  \\
T2-E3 & $32\times 16^3$ & 1.185 & 4313 & 0.675170(50) & 2.99(39) & 0.03751(38) & 4.66(75)  \\
T2-E4 & $32\times 16^3$ & 1.19 & 5404 & 0.676637(44) & 3.29(40) & 0.02474(28) & 7.9(1.5)  \\
\hline
T2-F0 & $64\times 24^3$ & 1.18 & 458 & 0.673737(46) & 4.0(1.9) & 0.04436(51) & 3.5(1.5)  \\
T2-F1 & $64\times 24^3$ & 1.185 & 291 & 0.675184(59) & 2.3(1.1) & 0.02836(59) & 4.2(2.5)  \\
T2-F2 & $64\times 24^3$ & 1.19 & 349 & 0.676649(52) & 1.63(59) & 0.01520(39) & 5.7(3.6)  \\
\end{tabular}
\label{tab:T3runs}
\caption{List of runs for the $T2$ theory at inverse coupling $\beta=2.25$. In the last two columns we give the average of the smallest eigenvalue of $|Q^{eo}_m|$ and its integrated autocorrelation time.}
}
 
This theory allows us to perform a non-trivial check of our code
generation. The link variables in the higher representation can be
built in two different manners: in one case we construct real matrices
according to Eq.~(\ref{eq:adj}), while in the other case we use our
generic algorithm for constructing the $2S$ complex representation of
SU($N$), as described in App.~\ref{app:b}. As expected, for a given
configuration in the fundamental representation the two
representations yield exactly the same matrices; the generic algorithm
produces complex $3 \times 3$ matrices with vanishing imaginary
parts. Evolving the configuration using the HMC with the two different
representations yields compatible results as one can see by the
comparison of the runs T2-A1 and T2-A1$^\prime$ shown in
Fig.~\ref{fig:2Scheck} for the case of the plaquette.
%
\FIGURE{
 \centering
  \epsfig{file=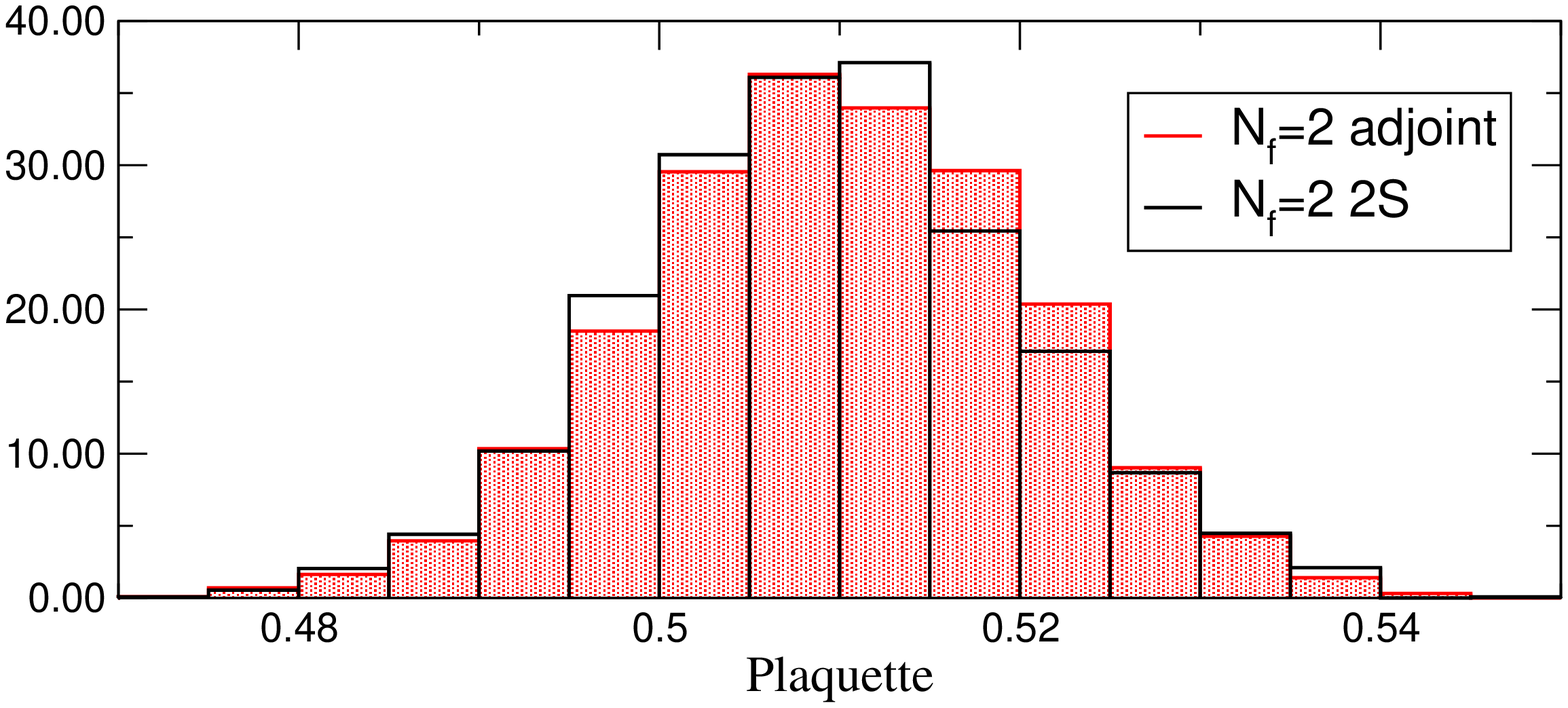,width=\columnwidth,clip}
  \label{fig:2Scheck}
  \caption{Normalized histogram of the plaquette distribution. 
    Data for the SU(2) theory with two
    flavours in the adjoint representation (red curve) are compared
    with data for two flavours in the two--index symmetric
    representation (black curve). }
}
%


The critical value of the hopping parameter has been computed in
perturbation theory in Ref.~\cite{DelDebbio:2008wb}, to which we refer
the reader for the notation and results used in what follows. Using
the cactus resummed formula for the theory under consideration, we
find:
\begin{equation}
\label{eq:2smc}
m_c=g_0^2\,2C_2(R)\times(-0.162857)/\tilde c_0
\end{equation}
where $\tilde c_0$ is a function of $g_0$ defined by solving the following equation:
\begin{equation}
\tilde c_0 = e^{-g_0^2/(16 \tilde c_0)} \left( 1 - \frac{g_0^2}{24 \tilde c_0} \right) \, .
\end{equation}
For $\beta=2.0$ and $\beta=2.25$ the above formula yields respectively $m_c=-1.73$ ($\kappa_c=0.220$) and $m_c=-1.47$ ($\kappa_c=0.198$). A non--perturbative determination of the chiral
limit in the space of bare parameters $(\beta,\kappa)$ can be obtained
by linear extrapolation of the PCAC mass, or of the mass squared of
the pseudoscalar Goldstone boson. Due to the power divergencies in the
renormalization of the bare quark mass, any perturbative computation
of $m_c$ is expected to receive non--perturbative corrections
$\mathcal O(e^{-1/g_0^2(a)}/a)$. Therefore the perturbative
computation can at best yield an indication of the location of the
critical point. The detailed study of these quantities is presented in
the next section.

In order to check that the range of $\kappa$ values does not lead to
exceptionally small values of the eigenvalues of the Dirac operator,
we have monitored the distribution of the smallest eigenvalue of $Q_m^{eo}$
following the study in Ref.~\cite{DelDebbio:2005qa}.
%
\FIGURE{
  \epsfig{file=fig7.eps,width=\columnwidth,clip}
  \label{fig:su2adjeig}
  \caption{Probability distribution of the lowest eigenvalue of the
    Dirac operator for $\beta=2.25$. }
}
%
The histograms that describe the distribution of the lowest eigenvalue
of $|Q_m^{eo}|$ are displayed in Fig.~\ref{fig:su2adjeig} for the simulations
at $\beta=2.25$. At the values of $\kappa$ that we have considered, the
spectrum of the Dirac operator has a clear gap. This is reflected in a
smooth behaviour of the simulations even at the lightest masses, as
can be seen in the time--histories for the plaquette and the solver
number reported in Fig.~\ref{fig:2Stherm}. The results obtained with
fundamental fermions in Ref.~\cite{DelDebbio:2005qa} suggest that the
width of the distribution scales like $a/\sqrt{V}$ as the continuum
and thermodynamic limits are approached. A comparison of the standard deviations
of the eigenvalue distributions for the simulations at $\beta=2.25$ shows
that this scaling is well verified for the theory under
consideration (Fig.~\ref{fig:loweig_stdev}).
%
\FIGURE{
\centering
  \epsfig{file=fig8.eps,width=\columnwidth,clip}
  \label{fig:2Stherm}
  \caption{Time-histories for the plaquette and the solver number for the T2-B11 lattice at $\beta=2$.}
}
%
\FIGURE{
\centering
  \epsfig{file=fig9.eps,width=\columnwidth,clip}
  \label{fig:loweig_stdev}
  \caption{The distribution width $\Delta \lambda_0$ of the lowest eigenvalue of the Dirac operator scales like $V^{-1/2}$. Results for $\beta=2.25$.}
}

As a final test of the algorithm we monitored the MC integration forces for the RHMC algorithm.
Compatibly with the absence of exceptionally small eignvalues, these forces appear to behave smoothly as a function of the bare mass and we observe a very mild (consistent with zero) dependence on the size of the lattice at constant $am_0$. In the upper panel of Fig.~\ref{fig:forcevol} we show the modulus of the force $|F^a_F(x,\mu)|$ averaged over all the lattice points $x$ and all directions $\mu$ for each term of the rational approximation indexed by $n$. Data for two different volumes at the same value of the bare quark mass are shown corresponding to the lattices T2-E2 and T2-F0. The average fermionic force appears to change very little once the equilibrium is reached which is reflected in the plot by the fact that errors on this quantity is really small and not visible on the scale of the graph. The realitive error for the average force ranges from a few permille for $n=1$ to less than one percent for $n=11$. The variation with volume is also very small: the two average values appear to be always consistent with each other within errors. 
In the bottom panel of Fig.~\ref{fig:forcevol} we report for completeness the size of the rational approximation coefficients used in the simulation. Their variation is also quite small, given that the distribution of the smallest and largest eigenvalue of the Dirac operator is quite narrow as discussed above.

The dependence of the average fermionic forces on the bare quark mass is shown in Fig.~\ref{fig:forcemass} for the T2-F lattice. The change with the quark mass is quite small but visible in this case, and more pronounced for higher values of $n$, corresponding to smaller shifts.  

\subsection{Autocorrelation times}
The integrated autocorrelation time for the plaquette during our runs
is computed in units of molecular dynamics time following
Ref.~\cite{Wolff:2003sm}. The results are reported in
Tab.~\ref{tab:T2runs}, where an increase of the autocorrelation time
is clearly visible as we move to lighter fermion masses. In all cases
the autocorrelation time is short compared to the total length of the
simulations, which guarantees that the binned configurations used for
our analyses can be considered as independent and therefore the
statistical error is correctly estimated from the variance of the
ensembles. It is worthwhile to emphasize that extensive simulations of
QCD have shown that the autocorrelation time depends on the trajectory
length~\cite{Meyer:2006ty}, and on the lattice spacing. Our runs are
all performed with trajectories of unit length. Moreover a critical
slowing down of the topological modes has been highlighted in the
time--histories of the topological
charge~\cite{DelDebbio:2004xh}. Therefore we expect that the
observables that are more strongly correlated with topology, show a
greater correlation length as we approach the continuum limit. These
issues need to be kept in mind as the simulations evolve and more
precise estimates of the phenomenological quantities will be
sought. Finally, one should keep in mind that autocorrelations depend
on the observables, and should be monitored for all the relevant
quantities in order to fully control the systematics in the
simulations.  For our runs at $\beta=2.25$ we changed the MC
integrator parameters to keep the plaquette autocorrelation time
roughly constant, which is why the data in Tab.~\ref{tab:T3runs} do
not show the same increase in autocorrelation times.
\FIGURE{
\centering
  \epsfig{file=fig10.eps,width=\columnwidth,clip}
  \label{fig:forcevol}
  \caption{Volume dependence of the average MC forces for each term in the rational approximation (upper panel) and average value of the approximation coefficients (lower panel). Data are obtained on the T2-E2 and T2-F0 lattices. Error bars in all cases are much smaller than the size of the symbols.}
}
\FIGURE{
\centering
  \epsfig{file=fig11.eps,width=\columnwidth,clip}
  \label{fig:forcemass}
  \caption{Mass dependence of the average MC fermionic forces for each term in the rational approximation. Data are obtained on the T2-F lattices. Error bars are not shown as they are much smaller than the symbols used in the plot.}
}

\section{Benchmark mesonic observables}
\label{sec:pheno}

Phenomenological results on the spectrum of the T2 theory are
extracted from the study of lattice two--point functions.  In this
study we do not attempt to perform an investigation of the
phenomenology of this theory, which will be presented in a forthcoming
paper.  Our aim is to provide lattice results useful as a benchmark
for future investigations, and to define the observables and the
analysis procedure that we use here and in subsequent papers.  In this
section we describe in detail the procedure used to extract mesonic
observables, such as the mass of the pseuodoscalar or vector mesons,
and the corresponding results on small lattices at $\beta=2.0$.

\subsection{Definitions}
Let $\Gamma$ and $\Gamma^\prime$ be two generic matrices in the
Clifford algebra, we define the two--point correlator at zero momentum
as follows:
\begin{equation}
f_{\Gamma\Gamma^\prime}(t) = \sum_\bfx \langle \bar\psi_1(\bfx,t) \Gamma
\psi_2(\bfx,t) \, \bar\psi_2(0) \Gamma^\prime \psi_1(0) \rangle\, ,
\end{equation}
where $\psi_1$ and $\psi_2$ represent two different flavors of
degenerate fermion fields, so that we only consider flavor
non--singlet bilinears.  Denoting the space--time position $(\bfx,t)$
by $x$ and performing the Wick contractions yields:
\begin{eqnarray}
\langle \bar\psi_1(x) \Gamma
\psi_2(x) \, \bar\psi_2(0) \Gamma^\prime \psi_1(0) \rangle &=&
- \mathrm{tr} \left[ \Gamma S(x) \Gamma^\prime S(-x) \right] 
\nonumber \\
&=& - \mathrm{tr} \left[ \Gamma S(x) \Gamma^\prime \gamma_5 
S^\dagger(x) \gamma_5 \right] \nonumber 
\end{eqnarray}
In practice we invert the Hermitean Dirac operator $Q=\gamma_5 D$ by
solving the equation:
\begin{equation}
Q(x,y)_{AB} \eta^{\bar A,0}_B(y) = \delta_{A,\bar A} \delta_{x,0}
\end{equation}
where capital Latin letters like $A=\{a,\alpha\}$ are collective
indices for color and spin, and $\bar A$, $x=0$ is the position of
the source for the inverter. The inversion is performed using a QMR
recursive algorithm with even--odd preconditioning of the Dirac
operator, which is stopped when the residue is less than $10^{-8}$.
Using the solution $\eta$ obtained from the inversion, the above
correlator is re-expressed as:
\begin{equation}
\langle \ldots \rangle = - \tilde \Gamma_{AB} \eta^{C,y}_B(x)
\tilde \Gamma^\prime_{CD} \eta^{D,y}_A(x)^*
\end{equation}
where $\tilde \Gamma= \gamma_5 \Gamma$, and $\tilde \Gamma^\prime =
\gamma_5 \Gamma^\prime$.

Following Ref.~\cite{DelDebbio:2007pz}, masses and decay constants for
the pseudoscalar meson are extracted from the asymptotic behaviour of
the correlators $f_\mathrm{PP}$ and $f_\mathrm{AP}$ at large Euclidean
time. The pseudoscalar mass and the vacuum--to--meson matrix element
are obtained from the correlator of two pseudoscalar densities:
\begin{equation}
  \label{eq:PPasym}
  f_\mathrm{PP}(t) = -\frac{G^2_\mathrm{PS}}{M_\mathrm{PS}}
  \exp\left[-M_\mathrm{PS} t\right] + \ldots\ .
\end{equation}
The meson mass is obtained by fitting the effective mass to a
constant, while the coupling $G_\mathrm{PS}$ is extracted from the
amplitude of the two--point function $f_{PP}$. For the precise
definition of the effective mass and coupling, we refer the reader to
Ref.~\cite{DelDebbio:2007pz}. On the other hand the ratio
\begin{equation}
  \label{eq:PCACmass}
  m_\mathrm{eff}(t) = \frac12 \left[\left(\partial_0 + \partial_0^* \right)
    f_\mathrm{AP}(t)\right] / f_\mathrm{PP}(t)
\end{equation}
yields the PCAC mass $m$ with corrections of $\mathcal O(a)$ for the
unimproved theory. On the relatively small lattices that we have used
in this study, it is difficult to isolate clearly the contribution
from the lowest state, that dominates the large--time behaviour of
two--point correlators. Fig.~\ref{fig:pilat} illustrates the typical
quality of the plateau that is fitted to extract the pion mass. The
data on the plot correspond to the effective mass obtained from two
$f_{PP}$ two--point function, whose asymptotic behaviour is governed
by the mass of the lightest pseudoscalar meson.  Lattices with a
smaller time extent do not show such a clear plateau; for these
smaller lattices the determination of the masses is affected by larger
systematics due to the contamination from heavier states.
\FIGURE{
  \centering
  \epsfig{file=fig12.eps,width=\columnwidth,clip}  
  \caption{Effective mass plot for the pion mass. The 
    points correspond to the effective mass
    extracted from $f_{PP}(t)$. The data refer to
    the T2-F0 lattice.}
  \label{fig:pilat}
}
%

Note that the decay constant is not computed directly; it is obtained
from the values computed above as:
\begin{equation}
  \label{eq:FPSeq}
  F_\mathrm{PS}=\frac{m}{M_\mathrm{PS}^2} G_\mathrm{PS}.
\end{equation}
The decay constant extracted from bare lattice correlators is related
to its continuum counterpart by the renormalization constant $Z_A$
which has been recently computed in perturbation theory in
Ref.~\cite{DelDebbio:2008wb}.

Finally the mass of the vector state is extracted from the $f_\mathrm{VV}$
correlator, again following the procedure outlined in
Ref.~\cite{DelDebbio:2007pz}.

\subsection{Results}
A first set of results were obtained from runs on the T2-A
lattices. Our results are summarized in Tab.~\ref{tab:4lat}.
\TABLE{
  \centering
  \begin{tabular}{lccccc}
    lattice & $a m$ & $a M_\mathrm{PS}$ & $a^2 G_\mathrm{PS}$ & $a
    M_\mathrm{V}$ & $a F_\mathrm{PS}$\\
    \hline
T2-A1 & 1.0199(14) & 2.5711(12) & 1.134(8) & 2.5814(13) & 0.3499(11) \\
T2-A2 & 0.7997(14) & 2.2313(16) & 1.277(10) & 2.2516(17) & 0.4102(16) \\
T2-A3 & 0.6718(14) & 2.0212(19) & 1.351(12) & 2.0517(21) & 0.4442(20) \\
T2-A4 & 0.5188(13) & 1.7708(24) & 1.435(14) & 1.8222(26) & 0.4748(26) \\
T2-A5 & 0.3129(23) & 1.362(7)  & 1.356(34) & 1.457(8) & 0.4578(68) \\
T2-A6 & 0.2599(39) & 1.196(18) & 1.157(59) & 1.280(20)& 0.421(12) \\
T2-A7 & 0.1601(30) & 0.868(27) & 0.776(47) & 0.901(31)& 0.3301(97) \\ 
T2-A8 & 0.0921(16) & 0.677(24) & 0.607(27) & 0.684(28)& 0.2433(61) \\ 
    \hline
  \end{tabular}
  \caption{Results for the PCAC mass, for the pseudoscalar mass and
    coupling, for the vector meson mass and the pseudoscalar decay constant, 
		from simulations on the T2-A lattices. }
  \label{tab:4lat}
}
The results obtained on such small lattices are affected by large
systematic errors, and hence are not suitable for reliable
phenomenology. On one hand the limited extension of the T2-A lattices
in the time direction means that it is virtually impossible to
identify a proper plateau. As a consequence we estimate the mass of
the relevant states from the value of the effective mass at the centre
of the lattice.  For the heavier masses we find that our results are
in agreement with the data presented in
Ref.~\cite{Catterall:2007yx}\footnote{The reader may notice however
  that a different normalization for $a F_\mathrm{PS}$ has been used
  in Ref.~\cite{Catterall:2007yx}. The two choices differ by a factor
  $\sqrt{8}$.  Our choice for the normalization of $a F_\mathrm{PS}$
  yields $F_\pi=93~\mathrm{MeV}$ in QCD.  }. However, for these larger
values of the bare mass, we see that all masses are $\mathcal O(1)$ in
units of the inverse lattice spacing, and therefore we expect these
results to be affected by large lattice artefacts. Smaller masses are
needed in order to identify the chiral dynamics that is relevant for
phenomenological studies. The lattices T2-A6, T2-A7, and T2-A8 yield
smaller masses, but the lightest pseudoscalar meson has a mass $a
M_\mathrm{PS}=0.68$, which is still large in units of the UV
cutoff. The study of the eigenvalue distributions presented in the
previous section suggests that it is not possible to go to lighter
masses on the T2-A lattices without entering the regime where the
algorithm becomes unstable, or the system gets close to a phase
transition.

The results obtained on the T2-B lattices for the pseudoscalar mass,
the PCAC mass, the vector meson mass, the axial vector meson mass, and
the vacuum--to--meson matrix element are reported in
Tab.~\ref{tab:masses}. As a result of the reduced width in the
eigenvalue distribution, we can afford to simulate closer to the
chiral limit, with a PCAC mass smaller than 0.1.
%
\TABLE{
  \centering
  \begin{tabular}{lccccc}
    lattice & $a m$ & $a M_\mathrm{PS}$ & $a^2 G_\mathrm{PS}$ & $a
    M_\mathrm{V}$ & $M_\mathrm{V}/M_\mathrm{A}$ \\
    \hline
    T2-B7   & 0.2209(30) & 1.149(11) & 1.190(56) & 1.269(12) & 1.51(22) \\
    T2-B8   & 0.1874(40) & 1.044(18) & 1.032(73) & 1.163(20) & 1.75(26) \\
    T2-B9   & 0.1624(37) & 0.952(19) & 0.875(63) & 1.067(21) & 1.77(15) \\
    T2-B10  & 0.1307(42) & 0.838(29) & 0.714(66) & 0.952(30) & 1.50(18) \\
    T2-B11  & 0.0455(14) & 0.356(34) & 0.214(12) & 0.366(43) & 0.67(25) \\
    \hline
  \end{tabular}
  \caption{Fitted values for the masses and vacuum--to--meson matrix
    element in lattice units for the T2-B lattices.}
  \label{tab:masses}
}
%
\TABLE{
  \centering
  \begin{tabular}{lcccc}
    lattice & $a m$ & $a M_\mathrm{PS}^2/m$ & $a F_\mathrm{PS}$ &
    $M_\mathrm{V}/F_\mathrm{PS}$  \\
    \hline
    T2-B7   & 0.2209(30) & 
    5.98(14)           & 0.399(14)        & 3.19(12) \\
    T2-B8   & 0.1874(40) & 
    5.82(24)           & 0.354(20)        & 3.29(19) \\
    T2-B9   & 0.1624(37) & 
    5.58(26)           & 0.314(18)        & 3.40(20) \\
    T2-B10  & 0.1307(42) & 
    5.30(41)           & 0.266(21)        & 3.58(30) \\
    T2-B11  & 0.0455(14) & 
    2.79(39)           & 0.154(8)         & 2.38(31) \\
    \hline
  \end{tabular}
  \caption{Fitted values for the relevant combinations of $m$,
    $M_\mathrm{PS}$, $G_\mathrm{PS}$, and $M_\mathrm{V}$ for the T2-B
    lattices. The value of the PCAC mass is reported again for
    clarity.}
  \label{tab:combi}
}
%

Several combinations of interest are also summarized in
Tab.~\ref{tab:combi}: the ratio of the pseudoscalar mass squared to
the PCAC mass, the bare pseudoscalar decay constant, and the mass of
the vector meson in units of the pseudoscalar decay constant. They are
computed from the primary observables discussed above, and the error
propagation is done by the usual jackknife method. The main
phenomenological conclusions of this paper are obtained from these
values.

The computation of the pseudoscalar decay constant in the chiral limit
is the method of choice to set the lattice scale. In a technicolor
theory the decay constant is related to the vacuum expectation value
of the Higgs field, and therefore $F_\mathrm{PS}=250~\mathrm{GeV}$.  A
realistic determination of the physical value of $F_{PS}$ would be
beyond the scope of this work.  Here we simply report the values of
$F_{PS}$ in Tab.~\ref{tab:4lat} and \ref{tab:combi}.

The ratio $M_\mathrm{V}/F_\mathrm{PS}$ is also an interesting quantity
of a Technicolor theory. Our result for this quanity are summarized in
Tab.~\ref{tab:combi}.

We stress again that the results in this section are only benchmarks:
they are useful for checking future numerical simulations but not to
draw any solid phenomenological conclusion about this theory.

\section{Conclusions}
In this paper we have described the generalization of the HMC and RHMC
algorithms to the case of generic representations of the color group
SU($N$). We have developed a general framework that can deal with
Wilson fermions in arbitrary representations and generic number of
colors $N$. We have put the emphasis in describing in detail the
algorithmic issues that need to be faced in generalizing existing code
to arbitrary representations. Since simulations for these new theories
are still in their infancy, and even the simplest results are not
known a priori, numerous independent studies will be welcome.  We have
provided an extensive number of tests of the algorithm and a detailed
study of its behavior for the SU(2) gauge theory with fermions in the
adjoint representation.  Our benchmark result are ideally suited for
validating lattice simulations of theories with fermions in higher
representations.

We have presented benchmark results on small $8^3 \times 16$ lattices,
giving full the details of the procedure we used, so as to make all
the results presented in this work reproducible beyond any
ambiguity. On these small lattice, our results should be easily
reproducible without any major investement of computational resources.
We provided a large number of benchmark quantities, ranging from the
average value of the plaquette to mesonic observables.

The results in this paper are the first step in a more comprehensive
program that aims to study nonperturbative phenomena beyond
QCD. Robust results for the spectrum and decay constants in
technicolor theories are an important ingredient to search for
strongly--interacting BSM physics at the LHC. The techniques that have
been developed for QCD offer all the theoretical and algorithmic tools
needed to develop a comprehensive study of technicolor on the
lattice. Lattice results will be important in order to test these
theories as potential candidates for BSM physics. We stress again that
an uncompromising theoretical formulation is mandatory for such
studies that are investigating unknown theories in order to have some
real predictive power. First numerical results should be benchmarked
carefully against each other in order to verify that systematics are
under control. 

It is worthwhile to emphasize once again that the algorithmic
framework that we have developed here can be readily used to study the
planar orientifold equivalence and lattice supersymmetry. Preliminary
results in this direction have already appeared~\cite{Armoni:2008nq}.

\vspace{0.5truein}

\noindent
{\bf Acknowledgements}

\noindent
During the long gestation of this work, we have enjoyed discussing its
progress with many people. We want to thank Adi Armoni, Francis Bursa,
Simon Catterall, Biagio Lucini, Martin L\"uscher, Francesco Sannino,
Misha Shifman and Gabriele Veneziano for discussions about various
aspects of this work. LDD acknowledges the kind hospitality of the
Isaac Newton Institute, CERN, and Odense University while this work
was progressing. Workshops at the INI, the Niels Bohr Institute, and
Edinburgh provided a lively atmosphere for discussions. We thank SUPA
for funding the workshop in Edinburgh. LDD is funded by an STFC
Advanced Fellowship.  The work of C.P. has been supported by contract
No. DE-AC02-98CH10886 with the U.S. Department of Energy during the initial 
stages of this work. A.P. is supported by an STFC special project grant 
and by the ``Fondazione Angelo Della Riccia''.

\appendix

\section{Group--theoretical factors}
\label{app:a}
The normalization of the generators in a generic representation $R$
of $\mathrm{SU}(N)$ is fixed by requiring that:
\begin{equation}
[T^a_R, T^b_R] = {\rm i}\, f^{abc} T^c_R\,,
\end{equation}
where the structure constants $f^{abc}$ are the same in all representations.
We define:
\begin{eqnarray}
\mathrm{tr }_R \left(T^a T^b \right) = \mathrm{tr } \left(T^a_R T^b_R
\right) &=& T_R \delta^{ab}, \\
\sum_a \left(T^a_R T^a_R \right)_{AB} &=& C_2(R) \delta_{AB},
\end{eqnarray}
and hence: 
\begin{equation}
T_R = \frac{1}{N^2-1} C_2(R) d_R
\end{equation}
where $d_R$ is the dimension of the representation $R$.  The quadratic
Casimir operators may be computed from the Young tableaux of the
representation of $\mathrm{SU}(N)$ by using the formula:
\begin{equation}
C_2(R) =\frac{1}{2}\left(nN+ \sum_{i=1}^{m} n_i \left( n_i+1-2i
\right) - \frac{n^2}{N}\right)
\end{equation}
where $n$ is the number of boxes in the diagram, $i$ ranges over the
rows of the Young tableau, $m$ is the number of rows, and $n_i$ is the
number of boxes in the $i$-th row. 
%
\TABLE{
\centering
\begin{tabular}{r|c|c|c}
R    & $d_R$               & $T_R$           & $C_2(R)$            \\
\hline
fund & $N$                 & $1/2$       & $(N^2-1)/(2 N)$ \\        
Adj  & $N^2-1$             & $N$             & $N$ \\
2S   & $N(N+1)/2$ & $(N+2)/2$ & $C_2(F) 2(N+2)/(N+1)$ \\
2AS  & $N(N-1)/2$ & $(N-2)/2$ & $C_2(F) 2(N-2)/(N-1)$ \\
\hline
\end{tabular}
\caption{Group invariants used in this work}
\label{table1}
}

The quantities $d_R$, $T_R$, $C_2(R)$ are listed in Tab.~\ref{table1}
for the fundamental, adjoint, 2--index symmetric, and 2--index
antisymmetric representations.

\section{Two--index representations}
\label{app:b}

The hermitean generators $T^a_F$ for the fundamental representation
used are of the following form.  For each pair of integers $1\leq
k<l\leq N$, we define two generators as:
\begin{eqnarray}
(T^{ij,+}_F)_{mn}&=\frac12 (\delta_{mk}\delta_{nl}+\delta_{ml}\delta_{nk})\,\, ,\\
(T^{ij,-}_F)_{mn}&=\frac{i}2 (\delta_{mk}\delta_{nl}-\delta_{ml}\delta_{nk})\,\, ,
\end{eqnarray}
and for each $1\leq k<N$ one more generator is defined as:
\begin{eqnarray}
(T^k_F)=\frac1{\sqrt{2k(k+1)}} \mathrm{diag}(\underbrace{1,1,\ldots,-k}_{k+1\, \mathrm{terms}},0,\ldots,0)\,\, .
\end{eqnarray}

The generators $T_F^a$ are normalized so that $T_F=1/2$. 
The generators for the other representations are obtained as follows.

We first give the explicit form for the representation functions $R$
which map $U\rightarrow U^R$. We define for each representation an
orthonormal base $e_R$ for the appropriate vector space of matrices.

For the Adjoint representation we define the base $e_{Adj}$ for the
$N\times N$ traceless hermitean matrices to be
$e_{Adj}^a=T^a_F/\sqrt{T_F}$, $a=1,\dots,N^2-1$ (i.e. proportional to
the generators of the fundamental representation and normalized to 1.)

For the two-index Symmetric representation the base $e^{(ij)}_{S}$,
with $i\le j$, for the $N\times N$ symmetric matrices is given by:
\begin{eqnarray}
i\neq j \, ,\,\,\,&(e^{(ij)}_S)_{kl}&=\frac{1}{\sqrt{2}} (\delta_{ik}\delta_{jl}+\delta_{jk}\delta_{il} )\, , \\
i=j \, ,\,\,\,&(e^{(ii)}_S)_{kl}&=\delta_{ki}\delta_{li} \, .
\end{eqnarray}

For the two-index Antisymmetric representation the base
$e^{(ij)}_{AS}$, with $i<j$, for the $N\times N$ antisymmetric
matrices is given by:
\begin{equation}
(e^{(ij)}_S)_{kl}=\frac{1}{\sqrt{2}} (\delta_{ik}\delta_{jl}-\delta_{jk}\delta_{il} )\, .
\end{equation}

The maps $R$ are explicitly given by:
\begin{multline}
(R^{Adj} U)_{ab} = U^{Adj}_{ab} = \mathrm{tr}\left[ e^a_{Adj} U e^b_{Adj} U^\dagger\right]\,\, , \\
a,b=1,\dots,N^2-1\, ,
\end{multline}
\begin{multline}
(R^{S} U)_{(ij)(lk)} = U^{S}_{(ij)(lk)} = \mathrm{tr}\left[ (e^{(ij)}_{S})^\dagger U e^{(lk)}_S U^T\right]\,\, ,\\
i\le j, l\le k\, ,
\end{multline}
\begin{multline}
(R^{A} U)_{(ij)(lk)} = U^{A}_{(ij)(lk)} = \mathrm{tr}\left[ (e^{(ij)}_{A})^\dagger U e^{(lk)}_A U^T\right]\,\, ,\\
i< j, l< k\, .
\end{multline}

The generators $T_R^a$ used are defined as the image of the generators
in the fundamental under the differential of the maps $R$ defined
above: $T^a_R = R_* T^a_F$.  Explicit expression can easily be worked
out form the definition above.  The invariants $T_R$ and $C_2(R)$ for
the generators defined in this way are given in Tab.~\ref{table1}.

\bibliography{paper_v4}

\end{document}